\documentclass[final,authoryear,5p]{elsarticle}

\usepackage{epsfig}

\usepackage{amssymb}

\usepackage[ps2pdf,%
a4paper=true,%
breaklinks=true,%
colorlinks=true,%
pdfauthor={A. Buzzoni et al.},%
pdftitle={Template for manuscripts in Advances in Space Research}%
]{hyperref}

\journal{Advances in Space Research}

\begin{document}

%%%%%%%%%%%%%%%%%%%%%%%%%%%%%%%%%%%%%%%%%%%%%%%%%%%%%%%%%%%%%%%%%%%%%%%%%%%%%
%% Frontmatter
\begin{frontmatter}

\title{Optical tracking of deep-space spacecraft in Halo L2 orbits and beyond:\\ 
the Gaia mission as a pilot case\tnoteref{footnote1}}
\tnotetext[footnote1]{Based on observations collected
at the Cassini Telescope of the Loiano Observatory, Italy}

\author{Alberto Buzzoni\corref{cor}}
%\author{Alberto Buzzoni}
\cortext[cor]{Corresponding author:}
\ead{alberto.buzzoni@oabo.inaf.it}
\author{Giuseppe Altavilla}
\author{Silvia Galleti}

\address{INAF - Osservatorio Astronomico di Bologna, Via Ranzani 1 40127 Bologna Italy}

\begin{abstract}
We tackle the problem of accurate optical tracking of distant man-made probes, on Halo orbit 
around the Earth-Sun libration point L2 and beyond, along interplanetary transfers. 
The improved performance of on-target tracking, especially when observing with small-class 
telescopes is assessed providing a general estimate of the expected S/N ratio in spacecraft 
detection.
The on-going {\sc Gaia} mission is taken as a pilot case for our analysis, reporting on fresh 
literature and original optical photometry and astrometric results. 

The probe has been located, along its projected nominal path, with quite high precision, within 
$0.13_{\pm 0.09}$~arcsec, or $0.9_{\pm0.6}$~km. Spacecraft color appears to be red, 
with $(V-R_c) = 1.1_{\pm0.2}$ and a bolometric correction to the $R_c$ band of 
$(Bol-R_c) = -1.1_{\pm0.2}$. The apparent magnitude, $R_c = 20.8_{\pm0.2}$, is much fainter 
than originally expected. These features lead
to suggest a lower limit for the Bond albedo $\alpha = 0.11_{\pm 0.05}$ and confirm that 
incident Sun light is strongly reddened by {\sc Gaia} through its on-board MLI 
blankets covering the solar shield.

Relying on the {\sc Gaia} figures, we found that VLT-class telescopes could yet be able to probe 
distant spacecraft heading Mars, up to 30 million km away, while a broader optical coverage of
the forthcoming missions to Venus and Mars could be envisaged, providing to deal with space 
vehicles of minimum effective area ${\cal A} \ge 10^6$~cm$^2$.
In addition to L2 surveys, 2m-class telescopes could also effectively flank standard 
radar-ranging techniques in deep-space probe tracking along Earth's gravity-assist maneuvers 
for interplanetary missions.
\end{abstract}


\end{frontmatter}

\parindent=0.5 cm

%%%%%%%%%%%%%%%%%%%%%%%%%%%%%%%%%%%%%%%%%%%%%%%%%%%%%%%%%%%%%%%%%%%%%%%%%%%%%
%% Main text

\begin{figure*}[th!]
\center
\includegraphics[width=0.58\hsize]{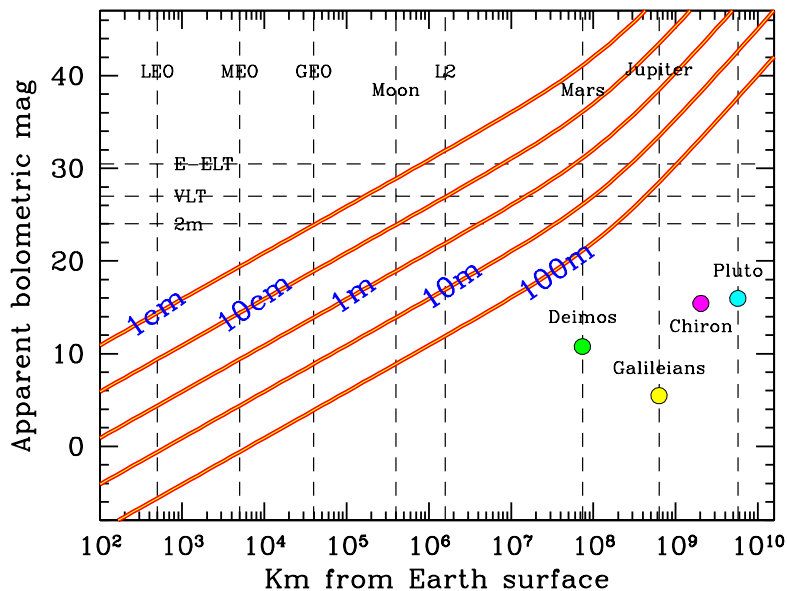}
\caption{The apparent bolometric magnitude for man-made spacecraft at increasing distance 
from Earth, according to eq.~(\ref{eq:2}). Probe scale-size is labelled along each
curve. The altitude of LEO (set to 500~km), MEO (5000~km) and GEO terrestrial orbits is marked, together
with a few small bodies in the Solar System and their reference interplanetary distances at Earth's 
opposition. The limiting magnitude reached by a 2m mid-class 
telescope, the 8m ESO VLT and the forthcoming 40m E-ELT telescope, when observing distant Sun-type 
stars, is also sketched on the plot. 
}
\label{f01}
\end{figure*}

\section{Introduction}
The exploitation of the Sun-Earth Lagrangian points, especially L1 and L2, along 
the Sun-Earth direction \citep{farquhar73,nariai75,rawal91,farquhar02} has been an 
extraordinary challenge for space exploration in the recent years. For their particular 
position, some $1.5~10^6$~km away, on opposite sides of Earth and therefore well beyond 
the Moon, both locations are ideal lookouts for astrophysical observatories aimed at studying 
the Sun (L1) and the deep Universe (L2), far from any anthropic contamination.
The L2 point, in particular, has been hosting a number of important astrophysical missions, 
starting with the {\sc Wmap}, {\sc Planck} and {\sc Herschel} probes, and currently continuing 
with the {\sc Gaia} mission, aimed at performing an exhaustive census of the Milky Way 
stellar population \citep{debruijne12,cacciari15}.
Following {\sc Gaia}, other major space facilities are planned to be located in a so-called Halo L2 orbit
in the forthcoming years. These include the James-Webb Space Telescope \citep[{\sc Jwst};][]{gardner06}, 
the {\sc Euclid} cosmological probe \citep{laureijs10}, and the {\sc Athena} X-ray observatory \citep{barret13}.

Optical ground tracking is yet of recognized importance for any L2 mission. For the co-rotating 
orbit to be maintained within its nominal figures, in fact, we need to carefully check spacecraft 
during its course along a complex Lissajous trajectory, as seen from Earth 
\citep[e.g.][]{bray67,zagouras85,liu07,kolemen12,dutt11,qiao14}. This is also of special interest
for any space observatory (like {\sc Gaia}, or the next {\sc Jwst}) as its absolute inertial position
is required with exquisite precision to allow, for instance, a confident measure of astronomical 
parallaxes of distant stars with the on-board instruments.
In this regard, radar-ranging techniques may actually provide a better measurement of spacecraft distance 
and radial velocity \citep{imbriale03}, but telescope observations, from their side, take 
advantage of a superior angular resolution, providing in principle more accurate 
astrometry and finer proper motion estimates \citep{altmann11}.

Compared to the observation of near-Earth satellites, however, optical tracking of distant probes, in L2 
and on route to even farther interplanetary distances, has to deal with much fainter target 
magnitudes, a drawback that urges a substantial improvement in terms of telescope skills and 
especially of observing techniques to effectively assess our deep-space situational awareness
\citep[e.g.][]{mooney06,ruprecht14,woods14}.

In this contribution we want therefore to briefly assess some technical issues (Sec.~2) dealing
with accurate ground tracking of deep-space probes at optical wavelength, taking fresh observations 
of the {\sc Gaia} spacecraft (Sec.~3) as a pilot case for tuning up our theoretical analysis. 
The relevant photometric figures for {\sc Gaia} will constrain the required telescope performance, 
for the optical observations to consistently complement standard radar-tracking techniques 
as in the forthcoming space missions to Mars and other planets of the solar system
(Sec.~4). Our conclusions will be briefly stressed in Sec.~5.

%__________________________________________________________________

\section{Apparent magnitude of distant spacecraft}
%__________________________________________________________________

Depending on its physical properties, a satellite under solar illumination reflects a fraction 
$\alpha$ (the so-called Bond albedo) of the incident flux. The remaining fraction of the input energy 
is retained and heats the body up to an equilibrium temperature that leads to a balance between 
the absorbed and re-emitted flux. At Earth's heliocentric distance, this temperature cannot 
exceed 120$^\circ$C \citep[e.g.][]{gilmore02}, so that thermal emission of spacecraft in the terrestrial 
neighborhood (and beyond) is only relevant at mid/far-infrared wavelength.

If a probe offers a cross-section $s^2$ to Sun's light, being $s$ its reference scale-size,
and if we assume the illuminated body to reflect isotropically, then the apparent bolometric magnitude of 
a spacecraft placed at a distance $d$ from Earth's surface (at Sun's opposition)\footnote{Although, 
strictly speaking, $d$ is a topocentric distance, to all extent, for a distant spacecraft in L2 
and beyond, it basically coincides with the geocentric distance, as well.} simply scales as the 
ratio of the incident solar flux at Earth and at the spacecraft distance, so that 
\begin{equation}
m_{\rm bol} - m^{\rm bol}_\odot = -2.5 \log \left[\left(\frac{D_\odot}{D_\odot+d}\right)^2
\left(\frac{s^2}{4 \pi d^2}\right)\right], 
\label{eq:1}
\end{equation}
where $D_\odot = 1.49\,10^{13}$~cm is the astronomical unit (AU) and
$m^{\rm bol}_\odot = -26.85$ is the apparent bolometric magnitude of the Sun, as seen from 
Earth \citep[e.g.][]{karttunen96}. 
With the relevant substitutions, and expressing the spacecraft distance $u = d/D_\odot$ in AU,
eq. (\ref{eq:1}) takes the form:
\begin{equation}
m_{\rm bol} = -5 \log s +5 \log [u\,(1+u)] +k,
\label{eq:2}
\end{equation}
where the numerical constant is 
\begin{equation}
k = 2.5 \log (4\pi D_\odot^2) +m^{\rm bol}_\odot = 41.77, 
\label{eq:2b}
\end{equation}
providing the satellite scale-size $s$ is set in cm.

In Fig.~\ref{f01} we report an illustrative summary of the expected bolometric magnitude for distant
man-made probes of different characteristic size, compared with a few small planetary bodies. Just as a 
guideline, the limiting magnitude as for observing Sun-type stars, reached by mid-class (2m aperture) and 
new-generation telescopes (i.e. the ESO 8m VLT and the forthcoming 40m E-ELT) is also marked on the 
plot. 

As a fraction $(1-\alpha)$ of the incident solar flux is ``diverted'' into the infrared, to convert 
the bolometric figures to other broad-band {\it optical} magnitudes, say for instance the 
Johnson-Cousins $R_c$ band, we have to dim the re-processed optical flux such as
\begin{equation}
m_R = m_{\rm bol} -BC_R' -2.5\,\log \alpha,
\label{eq:3}
\end{equation}
being $BC_R' = (m_{\rm bol} - m_R)$ the bolometric correction to the band, if satellite were a 
perfect acromatic reflector (i.e. for $\alpha \to 1$).
This value has to be estimated on the basis of satellite's physical and geometrical properties. 
If we lack this information, then either the solar value ($BC_R' = +0.3$)  
could be taken as a first approximation or, if a color is known for our target, the $BC_R'$ correction
for a more appropriate star could be chosen, as from standard calibrations in the literature 
\citep[e.g.][]{johnson66,bessell79,buzzoni10}.

\begin{figure}[!t]
\includegraphics[width=0.98\hsize]{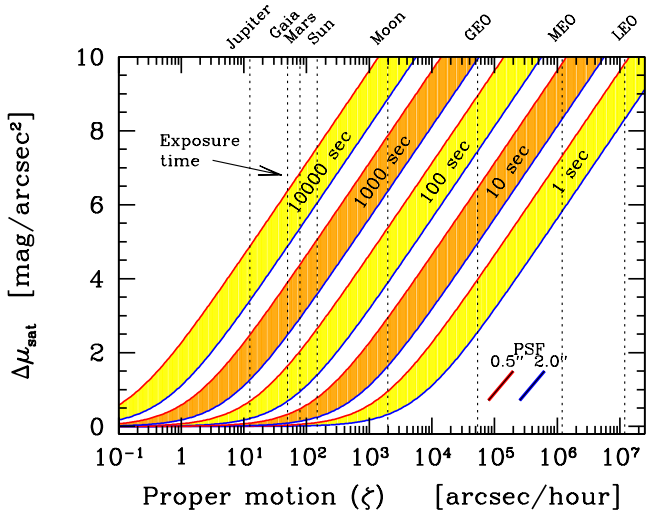}
\caption{The surface-brightness dimming for trailing satellites, according to 
eq.~(\ref{eq:dimming}). Different exposure times are assumed, as labelled. Each strip has a 
lower and upper envelope for a 2" and 0.5" FWHM seeing figure, 
respectively. The reference angular speed for satellites in LEO, MEO, and GEO \citep{veis63} is reported
together with the mean sky motion for other relevant solar bodies. In addition,
we also display the mean angular velocity of spacecraft {\sc Gaia}, along its Halo L2 
orbit. 
}
\label{f02}
 \end{figure}

%__________________________________________________________________

\subsection{Motion dimming}
%__________________________________________________________________

For fixed exposure time, any moving satellite tends to appear more elusive than fixed 
stars of the same magnitude on CCD images. 
In fact, while a star is supposed to concentrate most of its photons across a spot of area 
$a = \pi (\textsc{Fwhm}/2)^2$, whose FWHM depends on the telescope point-spread-function, a target moving with 
angular speed $\zeta$, along a time $t_{\rm exp}$, would spread its light across a larger area 
$a' = \zeta\,t_{\rm exp}\,\textsc{Fwhm} + \pi (\textsc{Fwhm}/2)^2$. 
If both the star and the moving target have the same magnitude, the latter would actually be harder to 
catch, because of a lower mean surface brightness and a correspondingly poorer S/N ratio. 
The surface-brightness dimming can be quantified in
\begin{equation}
\Delta \mu_{\rm sat}   = 2.5 \log (a'/a) = 2.5 \log \left(1 + {4\,t_{\rm exp}\,\zeta \over {\pi \textsc{Fwhm}}}\right).
\label{eq:dimming}
\end{equation}
Note that the effect does not depend on target magnitude.
Rather, and quite interestingly, it becomes more severe with increasing exposure time and with improving 
seeing (that is for a better FWHM value), as shown in Fig. \ref{f02}.

\begin{figure}[!t]
\includegraphics[width=\hsize]{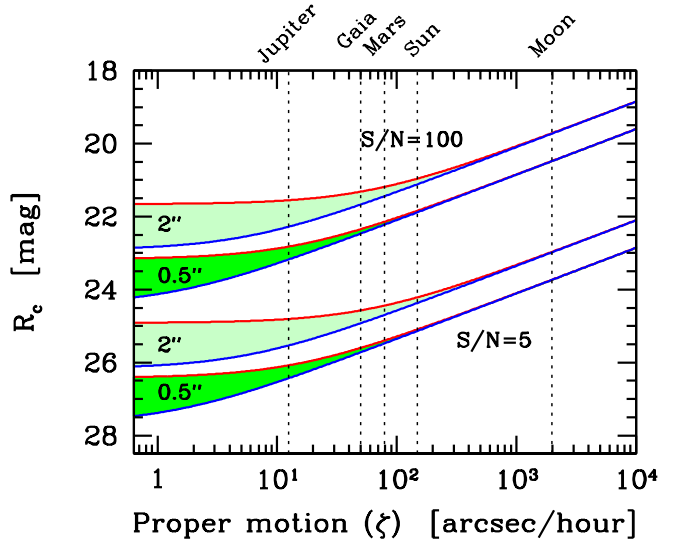}
\caption{The magnitude limit reached by a 8m telescope, with a $\textsc{Dqe} = 0.8$, at a 5 and 100 S/N 
detection level in 100 (upper envelope) and 1000 sec (lower envelope of the curves) exposure time, 
as from eq.~(\ref{eq:sn}). 
We assume to observe in the $R_c$ band under two extreme seeing conditions, namely 
with a FWHM of 0.5 and 2.0 arcsec, and with a dark sky ($\mu_{\rm sky}^R = 20.5$~mag arcsec$^{-2}$). 
The indicative proper motion of some reference objects is reported, as from Fig.~\ref{f02}.
}
\label{f03}
\end{figure}

According to telescope diameter $\cal{D}$ and the exposure time $t_{\rm exp}$, the S/N ratio
of a trailing object can eventually be computed as
\begin{equation}
\left({S\over N}\right)_{\rm trail} = {\cal R}\,{10^{-0.4\,\left[m_{\rm sat}-\left({{\mu_{\rm sky}-\Delta\mu_{\rm sat}}\over
2}\right)\right]}},
\label{eq:sn}
\end{equation}
with
\begin{equation}
{\cal R} = \frac{\cal{D}}{\textsc{Fwhm}} \left(n_o\,{\textsc{Dqe}}\,t_{\rm exp}\right)^{1/2}.
\label{eq:snb}
\end{equation}
In previous equations, $\mu_{\rm sky}$ is sky surface brightness, $n_o$ and {\sc Dqe} are the reference 
photon number for the magnitude zero point and the detector quantum efficiency, respectively, according 
to the photometric band of our observations. By definition, 
$n_o = (f_o\,\lambda_o\,\Delta_\lambda)/(h\,c)$, being $f_o$ the zero-mag reference flux, 
$\lambda_o$ and $\Delta_\lambda$ the effective wavelength and width of the photometric band, respectively,
$h$ the Planck constant and $c$ the speed of light, as usual.
For the $R_c$ band, $n_o = 1.05~10^{6}$ photons~cm$^{-2}$~s$^{-1}$ \citep{buzzoni05}. Notice that, 
if $\zeta \to 0$, then the $\Delta \mu_{\rm sat}$ term vanishes and eq. (\ref{eq:sn}) approaches 
the S/N ratio for a fixed star. 

As an illustrative case, in Fig.~\ref{f03} we report the limiting magnitude in the $R_c$ band, that 
can be achieved for a trailing target at a S/N~=~5 and 100 detection level with a 8m telescope 
($\textsc{Dqe} = 0.8$) in a 100 and 1000 sec exposure time. We assume to observe from a good astronomical 
site \citep[$\mu_{\rm sky}^R = 20.5$~mag arcsec$^{-2}$, e.g.][]{patat03}, under two extreme seeing conditions 
(namely 0.5" and 2.0" FWHM).
An important feature of the plot is that mag limit for high-speed satellites does not depend on exposure
time but it only improves with a better seeing or a bigger telescope. In the latter 
case, eq. (\ref{eq:sn}) (and the curves in Fig.~\ref{f03}) offsets by 
$\Delta R_{\rm lim} = 2.5 \log ({\cal D}/800)$, by expressing ${\cal D}$ in cm.

%__________________________________________________________________

\section{The {Gaia} observations as a benchmark}
%__________________________________________________________________

Previous arguments make clear that, when observing distant spacecraft, {\it on-target} 
telescope tracking, such as to compensate target motion, is the mandatory requirement
for letting any faint target literally  ``emerge'' from the background noise. 
Compared to the output of eq. (\ref{eq:sn}), in fact, the expected improvement in the S/N ratio
for the latter case is of the order of
\begin{equation}
\left({S\over N}\right)_{\rm track} = \left({S\over N}\right)_{\rm trail}  10^{+0.2\, \Delta \mu_{\rm sat}}.
\end{equation}

\begin{figure}[!t]
\includegraphics[width=0.46\hsize]{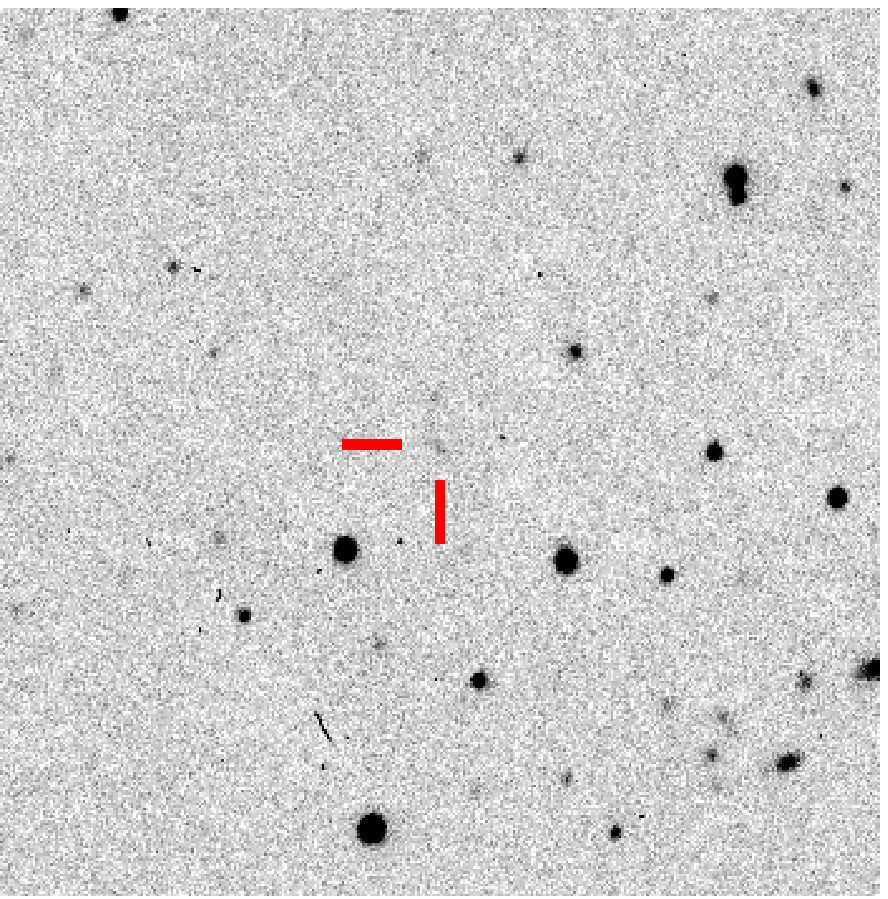}
\includegraphics[width=0.46\hsize]{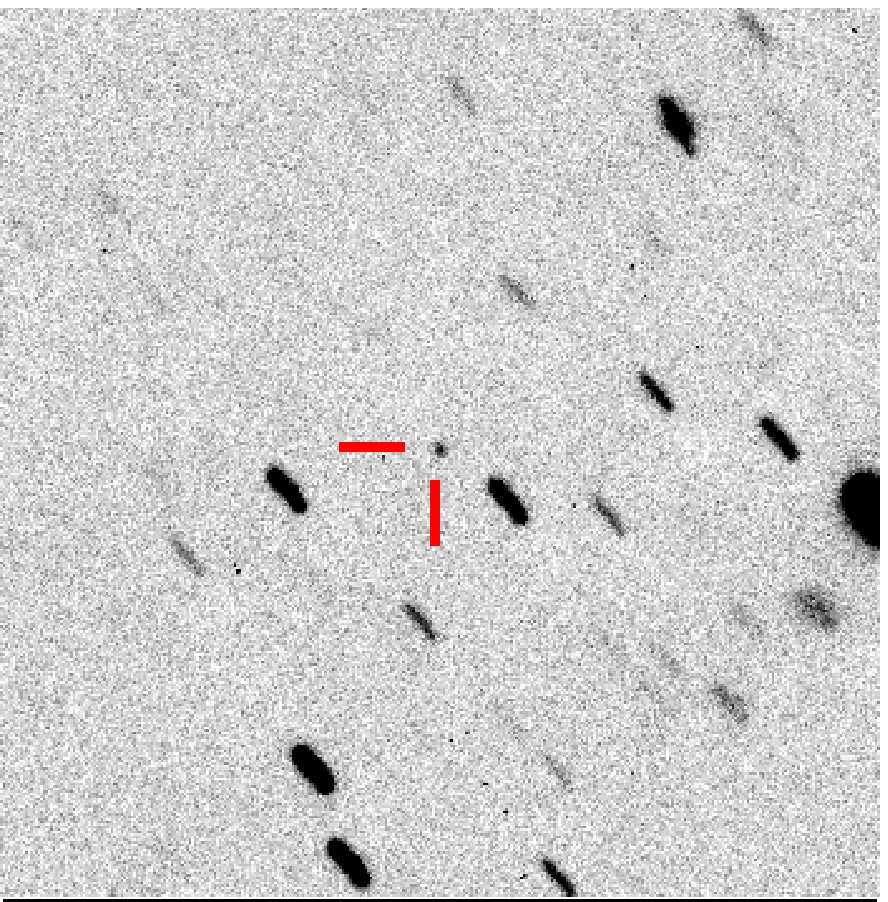}
\caption{The spacecraft {\sc Gaia}, as detected in the night of Oct 17-18, 2014 with the 1.52m
telescope of the Loiano Observatory. The two panels are consecutive 300 sec exposures 
in "white" light (i.e. CCD with no photometric filter) of the same field with
sideral (left) and differential (right) telescope tracking on. The displayed field is $3\times 3$ arcmin
across. North is up, East to the left. Pixel size is 0.58 arcsec. See text for a discussion.
}
\label{f04}
\end{figure}

As a striking example in this sense, we report here on two recent observing sessions on the {\sc Gaia} 
probe, along its Halo L2 orbit. Observations have been carried out with the Cassini 1.52m telescope 
of the Loiano Observatory (Italy), in a six-months interval along the night of Oct 17-18, 2014 and
on March 26-27, 2015. The telescope was equipped with the 
BFOSC camera, carrying a EEV $1300\times1340$ px coated and back-illuminated CCD, with a FOV of 
$12.6\times13.0$ arcmin, and a pixel scale of 0.58 arcsec px$^{-1}$. 

In addition to a set of eight
``white''-light frames (that is by exposing CCD with no photometric filter), the 2014 frames
also consisted of six $V$, $R_c$, and Gunn $z$ exposures. Overall, the spacecraft was tracked 
along five hours. The 2015 night was just devoted to a one-shot check of {\sc Gaia}'s position 
and apparent luminosity in the $R_c$ and $V$ bands.
Both nights had photometric conditions along the observation windows, with 
an R-band FWHM seeing figure about $1.5^{\prime\prime}$-$1.7^{\prime\prime}$.
The relevant data of our observations are summarized in Table~\ref{t01}. All frames were taken 
in ``on-target'' tracking mode, with the only exception of one image (as marked in the
table), tracked in sideral mode for the illustrative purpose of Fig.~\ref{f04}.

In the figure we actually report two consecutive 300 sec exposures of the same field, along the
Oct 2014 night, with sideral (left panel) and differential (right panel) telescope tracking on. 
Just at first glance, it is evident that {\sc Gaia} can barely be appreciated when trailing across the field, 
while it clearly stands out when concentrating its light as in the right panel.

\begin{table}
%\centering
\caption{The Oct 17-18, 2014 and March 26-27, 2015 Gaia observations}
\label{t01}
\scriptsize
\begin{tabular}{cccll}        % centered columns (4 columns)
\hline
\noalign{\smallskip}
UTC$^{(a)}$ & \multicolumn{2}{c}{$\qquad$RA$^{(b,c)}$\hfill (J2000)\hfill Dec$^{(b,c)}\qquad$}& Mag$^{(c,d)}$ & Exp.$^{(e)}$ \\
hh:mm:ss.s & hh:mm:ss.sss & dd:mm:ss.ss &  & sec\\
\noalign{\smallskip}
\hline
\noalign{\smallskip}
\multicolumn{5}{c}{Oct 17-18, 2014 observing session}\\
\noalign{\smallskip}
20:08:55.5 & 01:49:13.459$_{\pm009}$ & +15:30:11.54$_{\pm43}$ & 20.6$_{\pm3}$ w$_R$  & 120 d \\
20:15:45.6 & 01:49:14.134$_{\pm049}$ & +15:30:20.50$_{\pm62}$ & 20.7$_{\pm3}$ w$_R$  & 300 d \\
20:23:56.1 & 01:49:14.951$_{\pm015}$ & +15:30:31.71$_{\pm21}$ & 21.1$_{\pm3}$ w$_R$  & 300 d \\
20:28:55.5 & 01:49:15.434$_{\pm028}$ & +15:30:38.12$_{\pm29}$ & 20.9$_{\pm3}$ w$_R$  & 300 d \\
20:34:08.3 & 01:49:15.907$_{\pm016}$ & +15:30:45.32$_{\pm21}$ & 20.6$_{\pm3}$ w$_R$  & 300 d \\
20:44:07.5 & 01:49:16.758$_{\pm014}$ & +15:30:57.88$_{\pm13}$ & 20.9$_{\pm3}$ w$_R$  & 300 d \\
20:55:32.7 & 01:49:17.695$_{\pm015}$ & +15:31:11.71$_{\pm14}$ & 20.5$_{\pm3}$ w$_R$  & 300 s \\
21:13:16.1 & 01:49:18.975$_{\pm017}$ & +15:31:34.35$_{\pm21}$ & 21.0$_{\pm2}$ R$_c$  & 300 d \\
21:29:09.9 & 01:49:20.007$_{\pm016}$ & +15:31:52.66$_{\pm16}$ & 22.2$_{\pm2}$ V      & 420 d \\
21:35:46.8 & 01:49:20.363$_{\pm016}$ & +15:32:00.45$_{\pm54}$ & 20.6$_{\pm4}$ z$_R$  & 300 d \\
21:44:25.3 & 01:49:20.896$_{\pm015}$ & +15:32:09.63$_{\pm09}$ & 20.7$_{\pm2}$ R$_c$  & 300 d \\
00:50:46.7 & 01:49:29.027$_{\pm005}$ & +15:34:37.50$_{\pm34}$ & 21.0$_{\pm2}$ R$_c$  & 300 d \\
00:58:26.2 & 01:49:29.475$_{\pm011}$ & +15:34:40.94$_{\pm25}$ & 20.8$_{\pm3}$ w$_R$  & 300 d \\
01:10:54.5 & 01:49:30.163$_{\pm008}$ & +15:34:45.55$_{\pm27}$ & 21.9$_{\pm2}$ V      & 600 d \\
\hline
\noalign{\smallskip}
\multicolumn{5}{c}{March 26-27, 2015 observing session}\\
\noalign{\smallskip}
00:00:14.5 & 12:28:27.063$_{\pm013}$ & +04:09:23.99$_{\pm20}$  & 20.7$_{\pm3}$ R$_c$ & 480 d \\
00:13:46.8 & 12:28:27.499$_{\pm013}$ & +04:09:02.05$_{\pm20}$  & 21.7$_{\pm3}$ V     & 900 d \\
\noalign{\smallskip}
\hline
\noalign{\smallskip}
\multicolumn{5}{l}{$^{(a)}$ Luminosity-weighted time barycenter, according to eq.~(\ref{eq:9})}\\
\multicolumn{5}{l}{$^{(b)}$ Assumed topocentric coordinates of the telescope:}\\ 
\multicolumn{5}{l}{~~~~~($\lambda,\phi,h$) = ($11^o20^{m}02.7^{s}\, E,$ $44^o15^\prime33.2^{\prime\prime}\, N, 745.3$~m a.s.l.)}\\ 
\multicolumn{5}{l}{$^{(c)}$ Error figures apply to the last digits of each entry}\\
\multicolumn{5}{l}{$^{(d)}$ Photometric bands: Johnson-Cousins bands (V,R$_c$),}\\
\multicolumn{5}{l}{~~~~~ R-converted white light (w$_R$), and Gunn z band (z$_R$)}\\
\multicolumn{5}{l}{$^{(e)}$ Tracking mode: d = differential (``on-target''), s = sideral}
\end{tabular}
\end{table}

%__________________________________________________________________

\subsection{Astrometry}
%__________________________________________________________________

Although limited, our observed database allowed us to benchamark the astrometric 
and photometric reduction procedure, in order to assess the realistic performance in 
spatially locating the spacecraft at such large distance from Earth and characterize 
its apparent photometric properties. A first crucial piece of information deals with 
astrometry.

For the astrometric solution of the CCD images, after standard reduction
procedures with {\sc IRAF}\footnote{See the URLs: {\sl http://iraf.noao.edu.}}, 
we relied on the {\tt WCSTools} package\footnote{See the URLs: {\sl http://tdc-www.harvard.edu/wcstools.}} 
to pick up a reference grid of template stars across the field 
of view and finally set the World Coordinate System (WCS) of the corresponding image.
The 50 brightest objects in each field, as detected by the {\sc IRAF} task {\tt STARFIND},
have been matched by the {\tt WCSTools IMWCS} task with the corresponding HST Guide 
star Catalogue II (GSC-II) \citep{russell90}\footnote{See the URL: 
{\it http://tdc-www.harward.edu/software/catalogues/\ gsc2.html}.} 
available by default to constrain the WCS across the frame.
The WCS is the relationship between the pixel coordinates and the celestial coordinate, 
and it is written in a standard way in the FITS header. 
The precise astrometric solution is then refined by means of the {\tt WCSTools SCAT} 
and the {\tt IRAF CCFIND} and {\tt CCMAP} tasks. In particular, {\tt SCAT} retrieves 
the GSC-II catalogue of the field, while {\tt CCFIND} uses the image WCS to convert
the celestial coordinates into image pixel coordinates and refine the latter ones by
using a centroid algorithm.
The matched coordinates are finally used by {\tt CCMAP} to compute a new plate solution
in the RA and DEC domain by a low-order polynomial fit. The astrometric solution is 
directly inspected by over-plotting the reference catalogue sources on the 
astrometrically calibrated image.

Overall, our procedure secured an internal astrometric accuracy of $0.38_{\pm 0.18}$~arcsec
(rms) in the photometric centroid determination of {\sc Gaia}'s {\it individual} observations 
(see Table~\ref{t01}).\footnote{For the illustrative scope of our analysis, we relied here on 
the GSC-II catalog, a default reference for {\tt WCSTools} to set the astrometric solution. 
Other catalogs could, however, be implemented providing to set them in appropriate format. 
Of these, especially the PPMXL \citep{roeser10} and CMC-15 \citep{cmc15} compilations may prove to be 
viable alternatives to improve accuracy at least in selected sky regions.}

Though slightly elongated, the star figures in the telescope images did not prove to severely 
affect the astrometric solution across the field, provided the trailing intensity of the
astrometric calibrators ($I$) does not vary (or just vary in a predictable way) 
along the exposure time.\footnote{Among others, relevant change along trailing intensity of stars could 
either be due to any small drift in the sideral and ``on-target'' telescope tracking, or 
to any unperceived tiny cloud crossing the field during exposure etc.}
This is, actually, a crucial requirement for the photometric barycenter of stellar tracks to 
consistently tie to the {\it luminosity-weighted} time barycenter
($t_o$) of the exposure. By definition, for the latter, we have
\begin{equation}
\frac{t_o}{t_{\rm exp}} = \frac{\int \tau I(\tau) d\tau}{\int I(\tau) d\tau}\, 
\frac{1}{\int d\tau} \le 1.
\label{eq:9}
\end{equation}
In case $I(\tau) = {\rm const}$, this relationship delivers the obvious solution $t_o = t_{\rm exp}/2$, 
so that the spacecraft position should be attributed to the mid-exposure time.

\begin{figure}
\includegraphics[width=0.9\hsize]{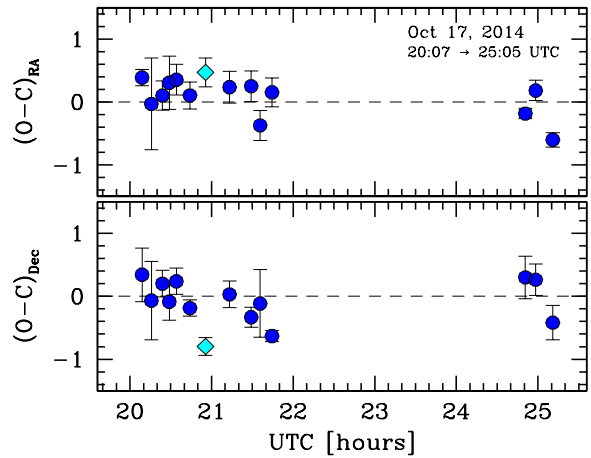}
\caption{The angular residuals of {\sc Gaia's} sky path along the night of Oct 17-18, 2014, as seen 
from the Loiano Observatory (UAI observatory code ``598''). The data of Table \ref{t01} are compared
with the corresponding JPL topocentric ephemeris. Residuals are in arcsec units, both for RA (upper panel)
and Dec (lower panel), in the sense ``Observed--Computed'', (O-C). The only sideral tracking observation
in our sample is singled out with a romb marker in both panels.  
}
\label{f05}
 \end{figure}

Clearly, the attainable astrometric resolution ($\theta_{\rm mas}$) also constrains the required 
accuracy, $\sigma(t_o)$ in setting $t_o$, depending on the spacecraft proper motion:
\begin{equation}
\sigma(t_o) \simeq \frac{3.6\,\theta_{\rm mas}}{\zeta} \quad {\rm [sec]}.
\end{equation}
In the equation, $\theta_{\rm mas}$ is expressed in mas (milliarcsec) unit and $\zeta$ in 
arcsec hr$^{-1}$, as in Fig.~\ref{f02}.
With the {\sc Gaia} typical figures, $t_o$ has to be known within a few seconds, at most.
According to the proper-motion constraints of Fig.~\ref{f02}, this is also the accuracy level for 
tracking interplanetary spacecraft, while a better tuned clock ($\sigma(t_o) \ll 0.2$~sec) is
devised, on the contrary, when observing fast-moving ($\zeta \gg 10^3$) Earth- and Moon-orbiting satellites.

\subsubsection{Locating L2 spacecraft}

\begin{figure}[!t]
\includegraphics[width=0.87\hsize]{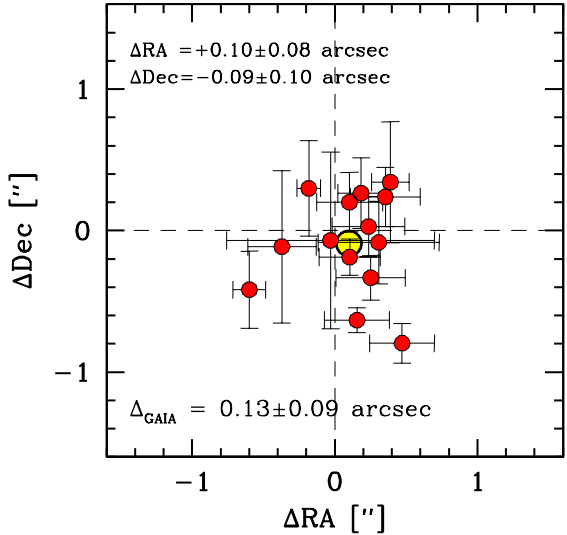}
\caption{Arcsec coordinate residuals (in the sense ``observed--computed'') of the {\sc Gaia} positions
along the night of Oct 17, 2014, with respect to the JPL topocentric ephemeris. 
Mean RA and Dec offsets, together with their 1-$\sigma$ uncertainty, are reported in the plot. 
When combined, these lead to a mean path offset with respect to the nominal figure of 
only $\Delta_{GAIA} = 0.13_{\pm 0.09}$~arcsec (big circle in the plot).
}
\label{f06}
\end{figure}

Figure~\ref{f05} shows that {\sc Gaia} position consistently compares with its 
nominal trajectory along the Oct 17, 2014 night, as 
predicted by the topocentric ephemeris from the JPL Horizons Solar System Dynamics (SSD)
Interface.\footnote{See the URL: {\sl http://ssd.jpl.nasa.gov/horizons.cgi.}}
The mean coordinate offsets of our data (in the sense ``observed--computed'') and their
corresponding $\sigma$ uncertainty amount to ($\Delta$RA, $\Delta$Dec) = ($0.10_{\pm 0.08}, 
-0.09_{\pm 0.10}$) arcsec, which lead to a mean transverse offset
$\Delta_{GAIA} = 0.13_{\pm 0.09}$~arcsec. Our observations demonstrate, therefore, that the 
spacecraft can correctly be located along its expected orbital figure with an angular accuracy 
$\theta_{mas}$ of the order of 90~mas (see Fig. \ref{f06}).\footnote{Quite consistently, 
notice that the derived 90~mas rms of the average coordinate residuals of 
Fig.~\ref{f06} fully compares with the $(380/\sqrt{N_{\rm obs}})\sim 100$~mas 
theoretical figure, as for the independent observations of Table~\ref{t01}.}
At the reported
distance of {\sc Gaia} along the night (namely $1.39\,10^6$ km, according to the JPL ephemeris), 
our measures point to a mean orbital offset of $0.9_{\pm0.6}$~km.

\begin{figure}[!b]
\includegraphics[width=0.87\hsize]{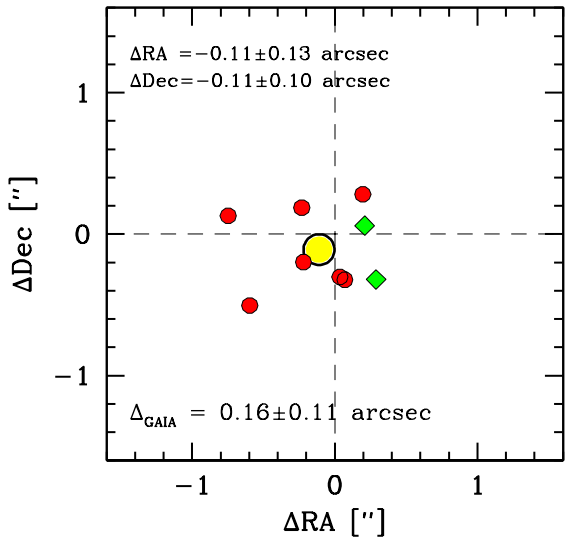}
\caption{Same as Fig.~\ref{f06} but for a set of astrometric measurements for the night of
Oct 14, 2014, taken with the 1.5m Mt. Lemmon (UAI code ``G96'') \citep[][dots]{kowalski14} 
and the 1.8m LPL/Spacewatch II (UAI code ``291'') telescopes \citep[][rombs]{tubbiolo14}.
Arcsec coordinate residuals are computed with respect to the correpsonding JPL topocentric ephemeris
leading to a mean orbital offset of $\Delta_{GAIA} = 0.16_{\pm 0.11}$~arcsec (big circle in the plot).
}
\label{f07}
\end{figure}

These figures nicely compare also with other almost parallel sets of   
observations, taken a few nights earlier (namely on Oct 14, 2014) by the 1.5m Mt.\ Lemmon 
\citep{kowalski14} and the 1.8m LPL Spacewatch~II \citep{tubbiolo14} telescopes. A summary of 
these data is shown in Fig.~\ref{f07}.
After rejecting one clear outlier in the LPL/ Spacewatch~II sample, they indicate, overall, a 
($\Delta$RA, $\Delta$Dec) = ($-0.11_{\pm 0.13}$, $-0.11_{\pm 0.10}$)~arcsec, leading to a mean
orbital offset of $\Delta_{GAIA} = 0.16_{\pm 0.11}$~arcsec (or $1.1_{\pm 0.7}$~km) with 
respect to the corresponding JPL ephemeris.

\begin{figure}[!t]
\includegraphics[width=0.87\hsize]{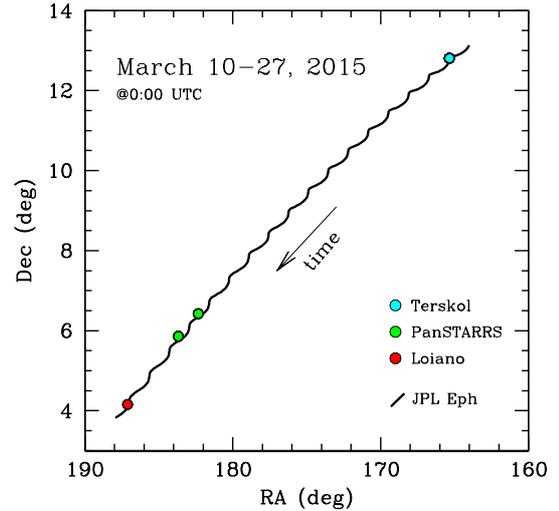}
\caption{The {\sc Gaia} sky path along the March 10-26, 2015 period.
The Terskol (UAI code ``B18'') \citep[][cyan marker]{velichko15}, PanSTARRS 
(UAI code ``F51'') \citep[][green dots]{gibson15} and Loiano (red dot) observations  are 
superposed to the JPL topocentric ephemeris (nominally for the B18 location, just as a guideline). 
Note the daily ``ripples'' of the {\sc Gaia} apparent orbit, due to the 
parallax effect of Earth rotation, that superposes to the overall Lissajous figure on larger scales.
}
\label{f09}
 \end{figure}

Closer to our 2015 observations, a supplementary set of 
10 astrometric measurements were provided by \citet{velichko15} in the night of March 10, 2015, at 
the 2m telescope of the Terskol Observatory (Russia), together with six estimates of \citet{gibson15}
with the 1.8m Pan-STARRS telescope in Haleakala (Hawaii, USA) on the nights of March 23 and 24, 2015 
(see Fig.~\ref{f09}).

\begin{figure}
\includegraphics[width=0.87\hsize]{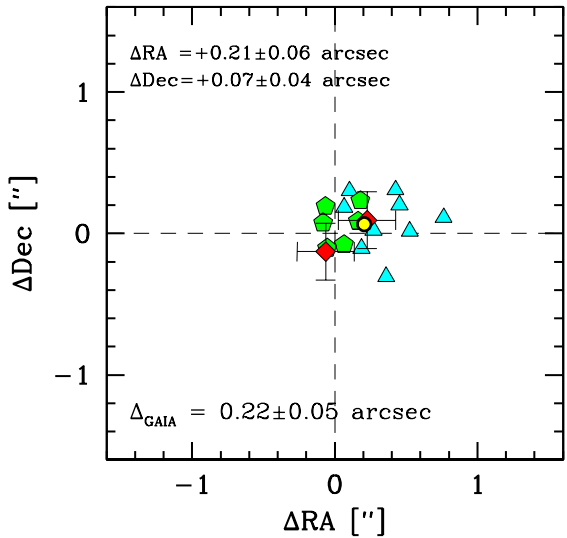}
\caption{Same as Fig.~\ref{f06} but for a coarser set of observations taken
during March 2015 at the Terskol 2m telescope (triangles), the 1.8m Pan-STARRS telescope (pentagons) 
and the Loiano 1.52m telescope (rombs, see Table~\ref{t01}).
Arcsec coordinate residuals are computed with respect to the appropriate JPL topocentric ephemeris
for each observatory. Error bars are only available for our observations, as from Table~\ref{t01}.
The merged set of data leads to a mean orbital offset of 
$\Delta_{GAIA} = 0.22_{\pm 0.05}$~arcsec (yellow circle in the plot) along the spanned period.
}
\label{f11}
\end{figure}

This coarser set of observations is merged in Fig.~\ref{f11} with our data and leads, overall, to
($\Delta$RA, $\Delta$Dec) = ($0.21_{\pm 0.06}$, $0.07_{\pm 0.04}$) arcsec, that is a mean 
orbital offset of $\Delta_{GAIA} = 0.22_{\pm 0.05}$ arcsec (i.e.\ $1.6_{\pm 0.4}$~km).

%__________________________________________________________________

\subsubsection{Locating deep-space spacecraft}
%__________________________________________________________________

The brief overview of the {\sc Gaia} case convincingly supports the preliminary results  
of the on-going tracking campaign carried on in the framework of the {\sc Gaia} mission 
plan \citep[see][for details]{altmann12},
and proves that optical measurements, even carried out with small-class telescopes and under 
quite standard observing conditions, can easily achieve a superior accuracy when compared 
to radar tracking, to locate distant spacecraft across the sky.
This better score is the obvious consequence of the fact that, for fixed angular resolving 
power, it must be
\begin{equation}
\left(\nu\, D\right)_{\rm radio} \simeq \left(\nu D\right)_{\rm opt}.
\end{equation}
Therefore, if a radar dish is set to operate at some $X$-band frequency 
\citep[say 8 GHz, as for the Deep Space Network antennas; ][]{imbriale03}, compared to the 
$\sim 5\,10^{14}$~Hz of a telescope observing in the $V$ band,
then a factor of $60,000$ larger antenna is required to achieve the same angular performance
of the optical instrument. This means that a 60~km-wide(!) antenna (or radio-interferometric baseline) 
is needed to offset the (diffraction-limited) performance of a 1m telescope.

Clearly, one could argue that any ground-based telescope is eventually seeing (and not diffraction)
limited, thus restraining its resolving power to a fraction of arcsec, at best. 
Another important issue also deals with the inherent astrometric accuracy of reference stars
to set the absolute astrometric grid. As for the HST Guide Star Catalog 
GSC-II used here, currently one of the most populated source of data across the sky for astrometric
studies, this limit turns to be about 300~mas 
\citep{lasker08,bucciarelli08}.\footnote{The GSC-II provides about 6 reference stars per 
square arcmin, down to $V \sim 20$, although the reported accuracy refers only to the 
brightest stars ($V\le 18.5$) eventually used in our study. This figure is expected to 
drastically improve (down to $\sim30\mu$as for $V \sim 15$ stars) in the 
forthcoming years, just relying on the {\sc Gaia} star survey \citep{debruijne12}.}

On the other hand, multiple measurements are possible of the target differential position 
with respect to a number of reference (astrometric) stars across the field of view, and this 
eventually leads to improve target astrometric accuracy 
by a factor $\sqrt{N_{\rm stars}}$. In addition, further improvements could be envisaged in 
case a series of independent observations could be collected. Definitely, our {\sc Gaia} 
benchmarking showed that an internal accuracy better than 100~mas 
is a fully attainable goal for standard observations carried out with 2m-class telescopes.

\begin{figure}[!t]
\includegraphics[width=\hsize]{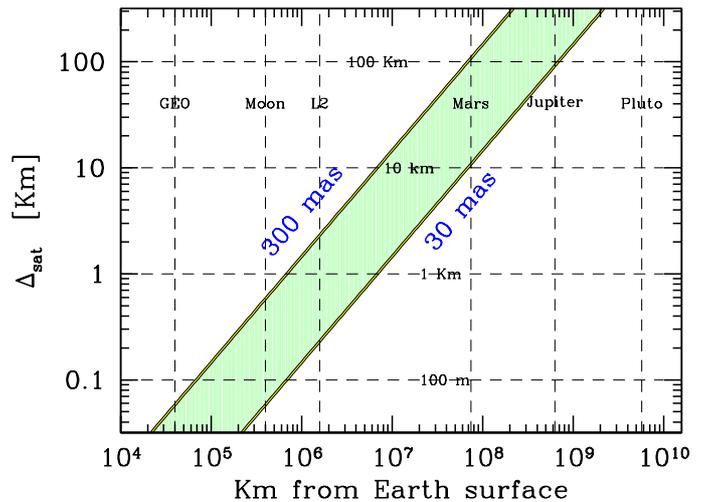}
\caption{The expected spatial resolution in detecting distant spacecraft within the
Solar System. The transverse component, projected on the sky, is assessed in terms of 
absolute resolution in kilometers at the different distances from Earth. Two representative
values for angular resolution of optical tracking are assumed, namely $\theta_{\mu as} = 300$~mas
and 30~mas, as labelled on the plot.
}
\label{f12}
\end{figure}

As far as even farther distances from Earth are considered, our arguments lead us to conclude 
that absolute position of man-made probes could be constrained within a nominal accuracy 
$\Delta_{\rm sat}$ of the order of 
\begin{equation}
\Delta_{\rm sat} = \frac{d_6\, \theta_{\rm mas}}{206}  \quad {\rm [km]},
\end{equation}
or
\begin{equation}
\Delta_{\rm sat} = 0.72\,u\, \theta_{\rm mas}  \quad {\rm [km]},
\end{equation}
assuming to express the geocentric distance in million km ($d_6$) or in AU ($u$, as in 
eq.~\ref{eq:2}), respectively. This relationship is displayed in Fig.~\ref{f12}, for values of 
$\theta_{\rm mas} = 30$ and 300~mas.

According to the figure, one can hope to optically pinpoint deep-space probes around Mars 
well within a few tens of km.
When complemented with radar information (more effective in ranging measurements and Doppler
velocity shift), the optical output could therefore be extremely valuable for an effective 
3D location of man-made probes at interplanetary distances.

Quite interestingly, Fig.~\ref{f12} also suggests that, under appropriate observing conditions 
(i.e.\ by masking Moon's overwhelming luminosity), Moon-orbiting spacecraft could be accurately 
located within a few hundred meters, a precision that could even raise to about 10m 
in the local framework, for instance when considering station-keeping maneuvering of Earth 
artificial satellites in geostationary orbit \citep{montojo11}.

%__________________________________________________________________

\subsection{Photometry}
%__________________________________________________________________

Both for the 2014 and 2015 observing sessions, photometry in $V$ and $R_c$ bands has been directly 
calibrated to standard magnitudes by observing two closeby \citet{landolt92} fields. Typical rms 
uncertainty for these observations is 0.2 mag, as summarized in Table \ref{t01}.
Although with a much broader passband, the effective wavelength of the CCD response curve 
happened to closely match that of the $R_c$ filter ($\lambda_o \simeq 6470$~\AA), and this eased 
a rough but still convenient transformation of the ``white'' instrumental magnitudes of the Oct 2014 
session into pseudo-equivalent $R_c$ figures (called $w_R$ in Table \ref{t01}). 
The whole conversion procedure relied on a grid of several 
field stars in common with the $R_c$ and ``white''-light images, and eventually led to a total 
error of 0.3 mag (rms) for the $w_R$ magnitudes.
Just for the sake of comparison, a similar procedure was also applied to the unique Gunn $z$ 
frame of the 2014 session, leading to a coarser $R_c$ proxy of  (i.e.\ $z_R$ in Table \ref{t01}), 
with a 0.4~mag overall error, given a large difference of the $z$ band effective wavelength 
($\lambda_o = 9040$~\AA). 

\begin{figure}[!t]
\includegraphics[width=\hsize]{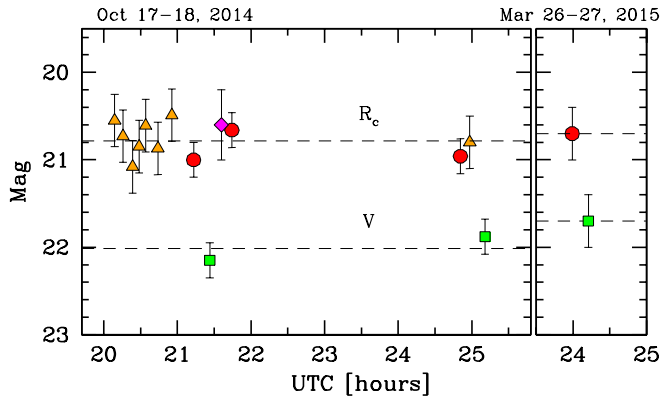}
\caption{The apparent magnitude of {\sc Gaia} in different photometric bands along the
Oct 2014 and March 2015 observing runs.
In addition to $R_c$ (dots) and $V$ (square) Johnson-Cousins bands, ``white''-light 
(triangles) and Gunn $z$ (romb marker) observations have been converted to the $R_c$ magnitude 
scale as discussed in the text. Dashed lines mark the mean $R_c$ and $V$ magnitude levels.
}
\label{f13}
\end{figure}

A plot of all the entries of Table \ref{t01} is shown in Fig. \ref{f13}, versus observing time.
Within the photometric accuracy of our observations, along the 2014 session {\sc Gaia} 
displayed a constant apparent luminosity with an average magnitude $\langle R_c \rangle = 20.8_{\pm0.2}$
and $\langle V \rangle = 22.0_{\pm0.2}$. These results are basically confirmed also by the
2015 observations, taken at the same topocentric distance and phase angle,\footnote{The phase 
angle is the angular distance between Sun (S) and Earth (E), as seen from the probe (P), that is 
$\phi = \widehat{\rm SPE}$. For our 2014 and 2015 observations $\phi \simeq 7^o$} which indicate
$\langle R_c \rangle = 20.7_{\pm0.2}$ and $\langle V \rangle = 21.7_{\pm0.2}$.
Overall, the observations point to a color of $(V-R_c) = 1.1_{\pm0.2}$, much redder than the 
solar value of $(V-R_c)_\odot = 0.4$, as estimated from \citet{pecaut13}.
By itself, this difference suggests that {\sc Gaia} seems to drastically re-process 
the incident solar flux. 

Furthermore, our results also point to a much fainter apparent magnitude 
than originally expected for the spacecraft \citep[namely $R_c \sim 17$-18, according to][]{altmann11},
based on the direct experience on previous L2 missions (especially {\sc Wmap} and {\sc Planck}).
Yet to a more updated analysis \citep{altmann14}, the reason of this discrepancy remains unclear.
Evidently, {\sc Gaia}'s faintness and its exceedingly red color have to be related to more inherent 
reflectance characteristic of the spacecraft structure, especially dealing with the extended Kapton 
multi-layer insulation (MLI) blankets that cover most of the sun-shield.

To help better investigate the problem, during the preliminary station-keeping 
maneuvers of Feb 27 and March 7, 2014, the probe has repeatedly been re-oriented, from its nominal
attitude configuration at solar aspect angle\footnote{This is the angle of the incident 
solar flux with respect to the normal of the satellite planar surface.} $\omega = 45^o$, such as
to have its sun-shield directly facing the Sun, that is with $\omega = 0^o$. 
This greatly brightened {\sc Gaia}'s apparent luminosity up to $R_c^{\rm peak} = 14.5_{\pm0.2}$
\citep{james14,jacques14,dupouy14,dupouy14b}.

The ``face-on'' experiment clearly indicated that {\sc Gaia}'s sun-shield reflectance displays 
an important directional pattern \citep[an effect often referred to as the ``opposition surge'', e.g.][]{warell04}
with spacecraft luminosity increasing by $\Delta {\rm mag} = 20.8-14.5 = 6.3_{\pm0.3}$~mag when 
changing $\omega$ from $45^o$ to $0^o$.
This figure has to be compared with the straight Lambertian prediction, that leads 
to a much shallower magnitude brightening of
\begin{equation}
\Delta {\rm mag} = -2.5 \log \left( \frac{\cos 45^o}{\cos 0^o} \right) \sim 0.4~{\rm mag}.
\label{eq:45deg}
\end{equation}

\subsubsection{Bolometric correction and spacecraft albedo}

According to eq.~(\ref{eq:2}), the expected bolometric magnitude of {\sc Gaia} at the relevant distance 
of our observations is $Bol = 16.9$~mag.\footnote{We adopt $s = 886$~cm as {\sc Gaia}'s relevant size
to be adopted in eq.~(\ref{eq:2}). See: {\sl http://www.esa.int/Our\_Activities/Space\_Science/\
Gaia/Gaia\_factsheet} for more details.}
As from eq.~(\ref{eq:3}), this figure can be linked to the apparent $R_c$ magnitude by correcting
for the appropriate bolometric correction ($BC^\prime_R$) and the spacecraft albedo ($\alpha$).
Both these quantities directly relates to the reflectance spectrum of the probe along the entire 
wavelength range, a feature that can only be poorly known {\it a priori} and needs to be conveniently
assessed ``on the fly''. 
For this task, as a part of the characterization checks during the early phases of the {\sc Gaia} mission 
\citep{altmann14}, narrow-band multicolor photometry was acquired in early 2014 with the {\sc Grond} 
imager at the ESO 2.2m telescope in La Silla. This allowed us to reconstruct the spacecraft reflectance 
spectrum along the 4500-13000~\AA\ wavelength range and therefrom derive the spacecraft colors
during the ``face-on'' experiments and later along the standard operation phase.

\begin{figure}[!t]
\includegraphics[width=0.97\hsize]{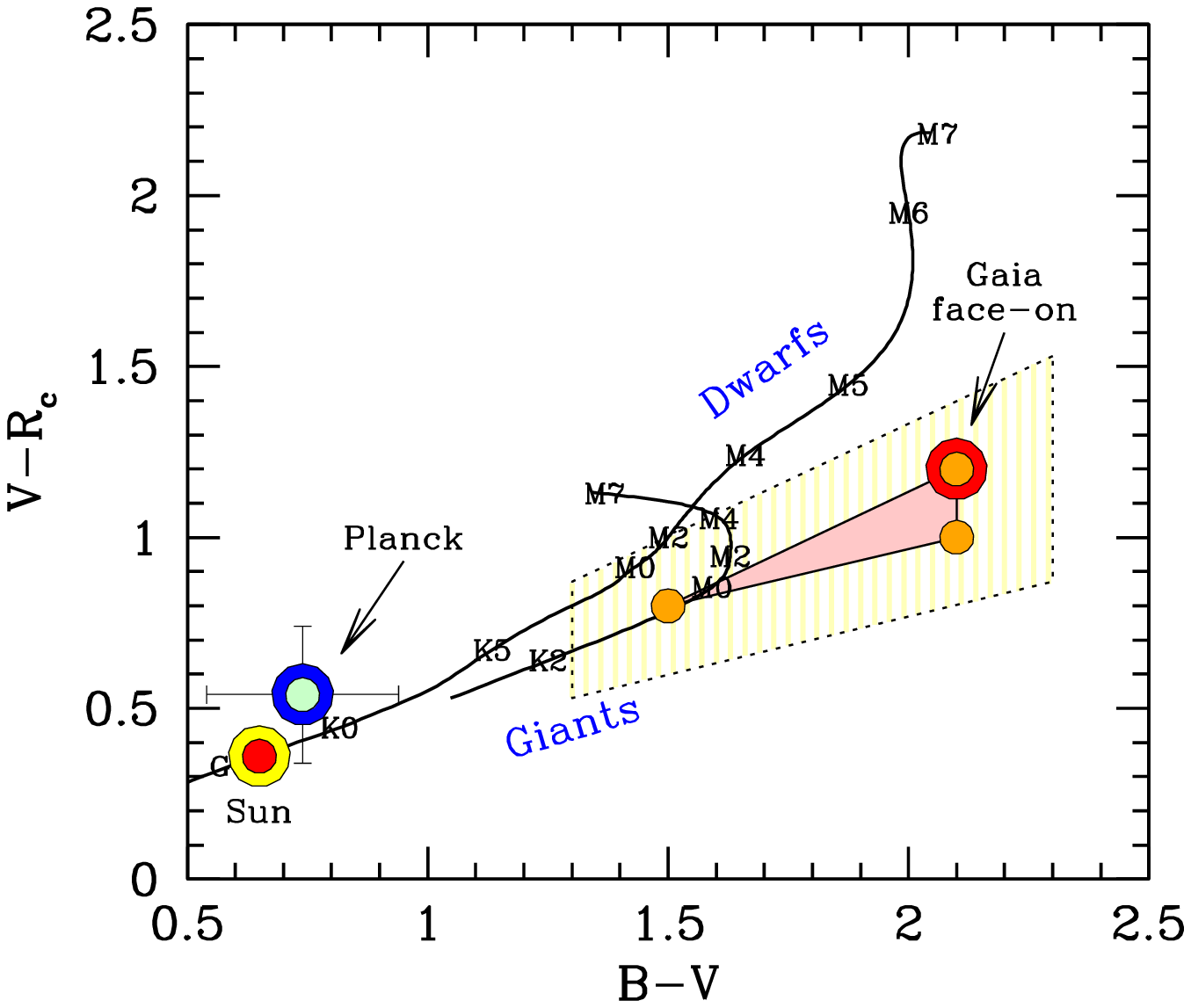}
\caption{The {\sc Gaia} color properties, as derived from the {\sc Grond} reflectance curves of 
\citet{altmann14}, are assessed in the $(B-V)$ vs.\ $(V-R_c)$ plane. The spacecraft location
is compared with the stellar locus for dwarf and giant stars of different spectral type (as
labelled on the plot), according to \citet{pecaut13} and
and \citet{houdashelt00}, respectively. The shaded triangular region on the plot edges
the allowed color range of {\sc Gaia} (including the experimented ``face-on'' orientation, as
discussed in the text), accounting for the reported variability of the spacecraft reflectance. 
As a reference, the Sun is also marked on the plot, together with the derived colors of the 
{\sc Planck} probe, according to \citet{altmann14}.
}
\label{f14}
\end{figure}

As far as the $(B-V)$ and the $(V-R_c)$ colors are concerned, the spacecraft 
location is compared in Fig.~\ref{f14} with the Sun and the locus for dwarf and giant stars of 
different spectral type. The empirical compilation of \citet{pecaut13} for MK-class V stars is 
adopted, for this comparison, together with the theoretical models of giant stars from \citet{houdashelt00}.
As a further interesting match, in the figure we also added the colors of {\sc Planck}, as derived from 
the original reflectance spectrum of \citet{altmann14}. Contrary to {\sc Planck}, which resulted to 
be a quite effective proxy of the Sun, {\sc Gaia} roughly behaves like a red dwarf of very late 
M spectral type, with an even more depleted $B$ luminosity (i.e.\ ``redder'' $B-V$ color).
In addition, one has to report from the {\sc Grond} data, an important (and yet partly unexplained) 
spectral variability of the probe, with its optical colors actually cornered within the wide region 
marked in Fig.~\ref{f14}. 

\begin{figure}[!t]
\includegraphics[width=0.97\hsize]{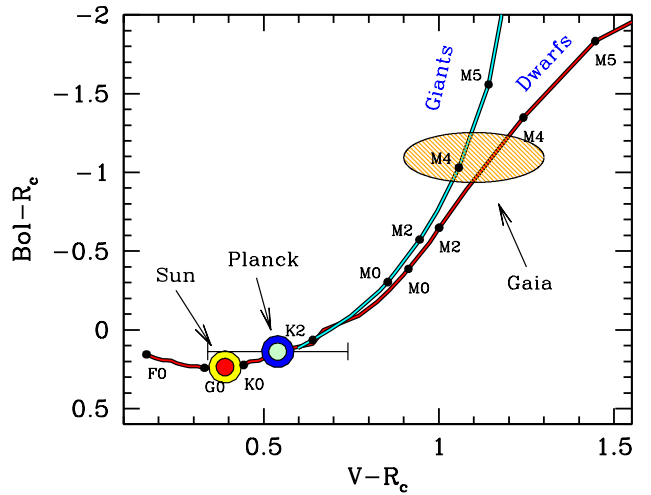}
\caption{The observed {\sc Gaia} $(V-R_c)$ color is contrasted in the plot to constrain the 
spacecraft bolometric correction. As for Fig.~\ref{f14}, the spacecraft location is compared 
with the stellar locus for dwarf and giant stars of different spectral type (as labelled on 
the plot), according to \citet{pecaut13} and and \citet{houdashelt00}, respectively, and 
with the relevant points for the Sun and {\sc Planck} probe.
}
\label{f15}
\end{figure}

On the basis of the reference $(V-R_c)$ color, in Fig.~\ref{f15} we try to envisage a suitable
bolometric correction by comparing, again, with the \citet{pecaut13} and \citet{houdashelt00}
stellar calibrations. Differently from {\sc Planck}, {\sc Gaia}'s color properties point to a much
larger correction, that we can tentatively place around a value of $BC^\prime_R \simeq -1.1\pm 0.2$~mag, 
although with a large uncertainty, further magnified by the spacecraft spectral variability.

With the relevant figures, under Lambertian assumptions, eq.~(\ref{eq:3}) takes the final form as
\begin{equation}
20.8_{\pm 0.2} -0.4_{\pm 0.3} = 16.9 +1.1_{\pm 0.2} -2.5\,\log \alpha,
\end{equation}
where the apparent $R_c$ magnitude in the l.h. term of the equation has been 
corrected\footnote{The error bar on the eq.~(\ref{eq:45deg}) magnitude correction accounts 
for the $\pm 15^o$ maximum phase angle excursion of the spacecraft along its libration orbit 
around L2, as seen from Earth.} to $\omega = 0^o$, according to eq.~(\ref{eq:45deg}).
This relationship eventually leads to a nominal value of the {\sc Gaia} albedo 
$\alpha = 0.11_{\pm 0.05}$.
 
Strictly speaking, however, this estimate is most likely to be taken as a {\it lower limit} 
for the true Bond albedo, due to sun-shield reflectance anisotropy, as we were 
discussing before.
In fact, by applying our arguments to the real ``face-on'' spacecraft configuration, 
one is led to conclude that {\sc Gaia} {\it could not be brighter} than 
$R_c^{\rm peak} \ge 16.9_{\pm 0.3} +1.1_{\pm 0.2} = 18_{\pm0.4}$, as originally estimated by 
\citet{altmann11}, but evidently at odds with the observed evidence. 

\begin{figure*}[t!]
\center
\includegraphics[width=0.85\hsize,clip=]{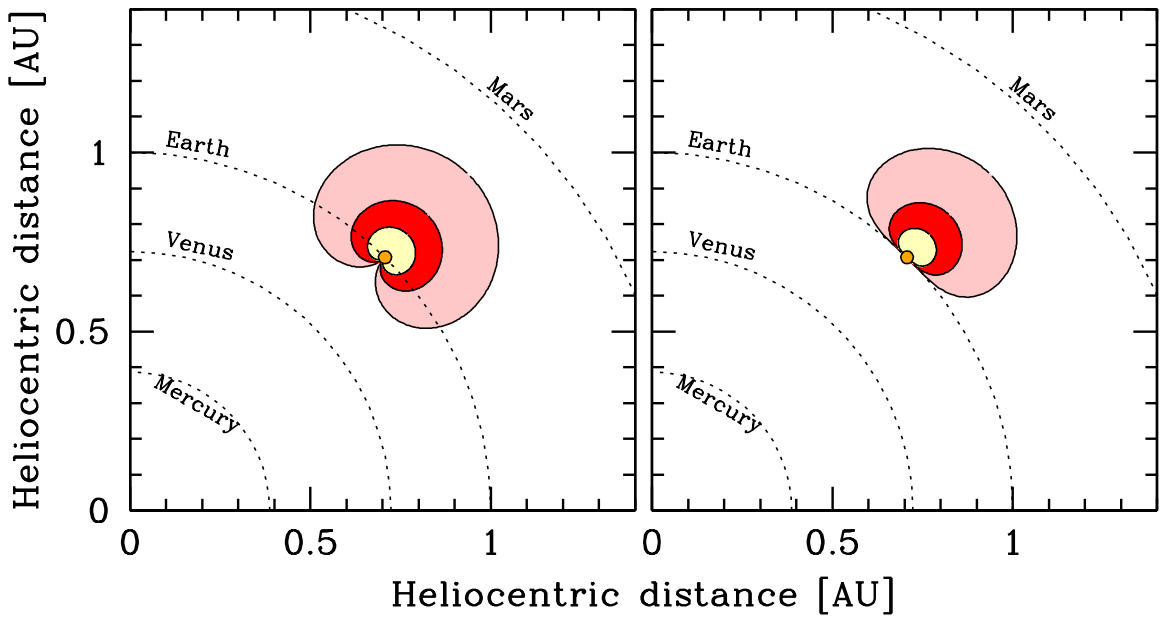}
\caption{Visibility maps of deep-space probes, as optically tracked from Earth with
1hr CCD exposure by different telescopes. 
Sky conditions assume a seeing $\textsc{Fwhm}= 1$~arcsec and 
$\mu_{\rm sky}^R = 20.5$~mag arcsec$^{-2}$.
The {\sc Gaia} effective area ${\cal A} = 10^5$~cm$^2$ (that is by assuming $\alpha = 0.11$ 
as a conservative estimate of the true Bond albedo) has been taken as a reference. 
The marked ``horizons'' in all panels assume to observe with a 2m (yellow region), 
8m (red), and 40m (pink) telescopes, at $(S/N)>5$ detection
threshold. The small orange dot marks Earth's influence sphere 
(edging the Lagrangian L1 and L2 points), throughout. 
Left panel refers to the case of a spherical spacecraft, while right panel
assumes a prevailing planar structure of the target. Mercury, Venus, Earth, and Mars
orbits are sketched, as a guideline.
}
\label{f16}
\end{figure*}

\section{Probing interplanetary distances}

When coupled with previous theoretical arguments, the {\sc Gaia} observations provide us 
with a useful tool to consistently size up the required telescope performance in order to 
detect deep-space probes at even larger distances from Earth, for instance when heading 
toward other planets of the solar system.
If we set a minimum $(S/N)$ threshold for target detection, then eq. (\ref{eq:sn}) 
(with ``on-target'' telescope tracking, that is for $\Delta \mu_{\rm sat} = 0$) 
can be solved to obtain the faintest possible magnitude we could reach and, by means of 
eq. (\ref{eq:3}) and (\ref{eq:2}), its corresponding maximum geocentric distance.

As probe needs to be under Sun's illumination to be detected from Earth, this also sets 
a constraint to the allowed range of the phase angle, depending on spacecraft 
(geocentric) distance and physical structure.
If planar components (i.e. large solar panels etc., either facing Earth or Sun) are the
prevailing features of the probe, then the simple geometrical arguments outlined in 
Sec.~3.2 for the case of {\sc Gaia} indicate that its apparent luminosity under different view 
angles scales as $\cos (\phi)$. This evidently implies that $|\phi| \le 90^o$ for the probe 
to be visible.
Much larger values of $\phi$ may, however, be allowed to roughly spherical shapes, as a 
cardioid-shaped luminosity law, as $[1+cos (\phi)]/2$, \citep[e.g.][]{meeus98} holds in 
this case.

The visibility map for the illustrative case of either a spheric or straight a planar 
satellite geometry is sketched in the two panels of Fig.~\ref{f16}, respectively. 
Calculations have been carried out for the {\sc Gaia}'s effective area 
${\cal A} = \alpha s^2 \simeq 10^5$~cm$^2$ (which assumes a conservative albedo value
$\alpha = 0.11$) supposing to detect the spacecraft at a $(S/N)$
ratio better than 5. With the 8m-class (or bigger) 
telescopes currently in use, one sees that an important region can be explored, well 
beyond the Earth's influence sphere (that embodies the Lagrangian L1 and L2 points)  
reaching, for instance, distant spacecraft along their initial Hohmann transfer 
to Mars, up to geocentric distances of some 30 million km (roughly 0.2~AU). 

Even 2m-class telescopes could however usefully track deep-space probes,
for instance during their flyby approach to Earth for gravity sling toward external planets. 
Successful observations of the Rosetta spacecraft \citep{glassmeier07}, along its 2005 and
2007 gravity-assist maneuvers \citep[e.g.][]{bedient05,dillon05,hill05,james05,juels05,
kusnirak05a,kusnirak05b,manca05,stevens05,ticha05,birtwhistle07a,birtwhistle07b,birtwhistle07c,
bittesini07a,bittesini07b,bittesini07c,donato07,hill07a,hill07b,kowalski07}\footnote{DASO Circulars
are made available in electronic form at the URL: {\sl http://www.minorplanetcenter.net/iau/DASO/DASO.html}.}
are a relevant example of such a kind of applications. The current Hayabusa~2 mission 
\citep{kuninaka13} could soon provide another intersting case to benchmark the method.

\begin{figure}
\includegraphics[width=0.9\hsize]{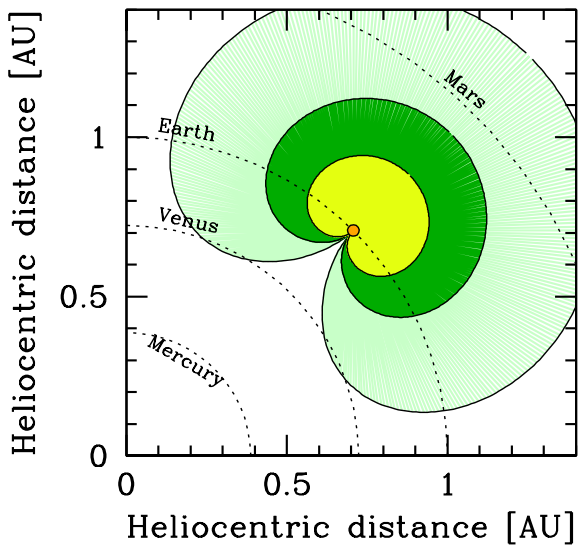}
\caption{Same as Fig.~\ref{f16} but for 2m (green), 8m (dark green) and 40m (pale green) 
telescopes looking (at a $S/N \sim 5$ or better threshold) at a deep-space spherical probe of  
{\sc Gaia}'s  10$\times$ enhanced effective area (namely ${\cal A} = 10^6$~cm$^2$). 
Under these more favourable circumstances, note that a VLT-class telescope could yet
confidently track a spacecraft along most of its Hohmann trajectory to Mars and Venus,
while a 2m telescope could, in general, effectively probe distant spacecraft 
some 40 million km away from Earth.
}
\label{f17}
\end{figure}

According to Fig.~\ref{f16}, optical tracking is of lesser point for any space
mission toward inner planets (i.e.\ Venus and Mercury). This is because the orbit
geometry leads probes usually to appear from Earth at large phase angles, thus
poorly reflecting Sun's light in our direction.
An enhanced spacecraft effective area-- either in terms of better reflectance 
to improve the albedo or directly by increasing the physical size of the
target, as likely the case of the forthcoming (manned) missions to Mars-- 
could be the obvious issue in this case,
as we show in the illustrative case of Fig.~\ref{f17}. By the way, a closer look to 
Fig.~\ref{f16} also shows that spherical spacecraft would better perform than other 
probes with more planar surfaces.

\section{Summary \& Conclusions}

We assessed in some detail the observing performance of optical telescope-tracking techniques, 
for accurate characterization of deep-space probes around the Earth-Sun libration point L2 and
beyond, toward interplanetary travels. We provided a general estimate of the expected S/N 
ratio, quantifying the superior improvement of on-target tracking, a mandatory requirement, 
especially when observing with small-class telescopes.

The current {\sc Gaia} mission has been taken as a pilot case for our discussion.
In this regard, we reported on fresh optical photometry and astrometric results, that led to
angularly locate the spacecraft across the sky within $0.13_{\pm 0.09}$~arcsec, or 
$0.9_{\pm0.6}$~km. Photometric results also indicate for 
{\sc Gaia} a quite red color, with $(V-R_c) = 1.1_{\pm0.2}$ and an apparent magnitude
$R_c = 20.8_{\pm0.2}$, much fainter than expected and dimmed by a large bolometric
correction $(Bol-R_c) = -1.1_{\pm0.2}$. These features lead
to a lower limit for the Bond albedo $\alpha = 0.11_{\pm 0.05}$ and confirm that 
the unabsorbed fraction of the incident Sun light is strongly reddened by
the MLI blankets, that cover the spacecraft sun-shield.

These observations provided us with the reference figures to consistently assess the detection
threshold for any deep-space probe toward inner and outer planets of the Solar System.
Waiting for next-generation (E-ELT) telescopes, yet VLT-class instruments could 
be able to track distant spacecraft, like {\sc Gaia}, along their initial Hohmann 
transfer to Mars, some 30 million km away. As successfully done with the Rosetta spacecraft,
we confirm that even 2m-class telescopes could usefully help track deep-space probes, 
along their Earth's gravity-assist maneuvers heading interplanetary targets.

We finally demonstrated that optical tracking from 8m ground telescopes could fully flank 
the standard radar-ranging techniques to probe distant spacecraft along the forthcoming 
missions to Venus and Mars, providing to deal with (round-shaped) space vehicles of minimum 
effective area ${\cal A} = \alpha s^2 \ge 10^6$~cm$^2$.

\section{Acknowledgements}
We would like to acknowledge the timely and very competent review of the two anonymous referees,
that greatly helped refine the original version of our study.

This work made extensive use of the {\sc IRAF} package, written and supported by the 
National Optical Astronomy Observatories (NOAO) in Tucson, Arizona. NOAO is operated by the 
Association of Universities for Research in Astronomy (AURA), Inc. under cooperative 
agreement with the National Science Foundation.

Our astrometric procedure made reference to the GSC-II Catalog, a joint project of the 
Space Telescope Science Institute (STScI) and the Osservatorio Astronomico di Torino (OATo), 
Italy. The STScI is operated by AURA under NASA contract NAS5-26555. The participation of OATo 
is supported by the Italian Council for Research in Astronomy. Additional support is provided 
by ESO and the Space Telescope European Coordinating Facility, the International GEMINI project 
and the European Space Agency (ESA).

Part of the data retrieval has been eased by the DEXTER graphic interface, hosted
at the German Astrophysical Virtual Observatory (GAVO) in Heidelberg.

We thank Roberto Gualandi and Italo Foppiani at OABo, Italy, for invaluable help along 
the telescope observations, and Jessica Mink at the Harvard-Smithsonian Center for 
Astrophysics of Cambridge MA, USA for her kind help to introduce us to the {\tt WCSTools} package.
%-------------------------------------------------------------------

\end{document}